\documentclass[twocolumn,showpacs,prl,aps]{revtex4}

\usepackage{epsfig}
\usepackage{amssymb,amsmath,amsthm}

\begin{document}
\title{Non-Gaussian statistics and extreme waves in a nonlinear optical cavity}

\author{A. Montina$^1$, U. Bortolozzo$^2$, S. Residori$^2$ and F. T. Arecchi$^{1,3}$} 

\address{
$^1$Dipartimento di Fisica, Universit\`a di Firenze, via Sansone 1, 50019 Sesto Fiorentino (FI), Italy
\\
$^2$INLN, Universit\'e de Nice Sophia-Antipolis, CNRS, 1361 route 
des Lucioles 06560 Valbonne, France  \\
$^3$INOA-CNR, largo E. Fermi 6, 50125 Firenze, Italy.
}

\date{\today}

\begin{abstract}
A unidirectional optical oscillator is built by using a liquid crystal 
light-valve that couples a pump beam with the modes of a nearly spherical 
cavity. For sufficiently high pump intensity, the cavity field
presents a complex spatio-temporal dynamics, accompanied by the emission 
of extreme waves and large deviations from the Gaussian statistics. 
We identify a mechanism of spatial symmetry breaking, due 
to a hypercycle-type amplification through the nonlocal coupling of 
the cavity field.
\end{abstract}


\maketitle

Extreme waves are anomalously large amplitude phenomena developing suddenly out 
of normal waves, living for a short time and appearing erratically with a small 
probability. These rare and extreme events have been observed since long time on 
the ocean surfaces \cite{oceanography}, and, in this context, have been called 
freak or rogue waves. Recently, rogue waves have been reported in an optical 
experiment \cite{Solli} and in acoustic turbulence in He II \cite{McClintock}. 
Associated with extreme waves are {\em L-shaped} statistics, with a probability of 
large peak occurrence much larger than predicted by Gaussian statistics.
Different mechanisms to explain the origin of extreme waves have been proposed, 
including nonlinear focusing via modulational instability (MI) \cite{Onorato} 
and focusing of line currents \cite{White}. 
In optical fibers, numerical simulations of the nonlinear Schr\"odinger 
equation (NLSE) \cite{Dudley}  have established a direct analogy between optical 
and water rogue waves. In a spatially extended system, the formation of large 
amplitude and localized pulses, so-called {\em optical needles}, has been evidenced 
by numerical simulations for transparent media with saturating self-focusing 
nonlinearity \cite{Rosanov}. Similar space-time phenomena, such as collapsing 
filaments, are also predicted in optical wave turbulence \cite{Dyachenko}. Recently, 
the spatio-temporal dynamics of MI-induced bright optical spots was observed in 
Ref.~\cite{Shih} and an algebraic power spectrum tail was reported due to 
momentum cascade \cite{Dylov}. Nevertheless, no experimental evidence has been 
given up to now of extreme waves in a spatially extended optical system.

Here, we report, what is, at our knowledge, the first experimental 
evidence of extreme waves and non-Gaussian statistics in a 2D spatially 
extended optical system. The experiment consists of a nonlinear optical cavity, 
formed by a unidirectional ring oscillator with a liquid crystal light-valve 
(LCLV) as the gain medium. While for low pump the amplitude follows a Gaussian 
statistics, for sufficiently high pump we observe large deviations from Gaussianity, 
accompanied by the emission of extreme waves that appear on the transverse 
profile of the optical beam as genuine spatiotemporal phenomena, developing 
erratically in time and in space. 
The observations are confirmed by numerical simulation of the full model equations.
Moreover, by introducing a mean-field simplified model, we show that extreme 
waves in the cavity are generated by a novel mechanism, which is based on a 
{\it hypercycle}-type amplification \cite{hypercycle} occurring via nonlocal 
coupling of different spatial regions.

\begin{figure}[h!]
\centerline{\includegraphics[width=\columnwidth]{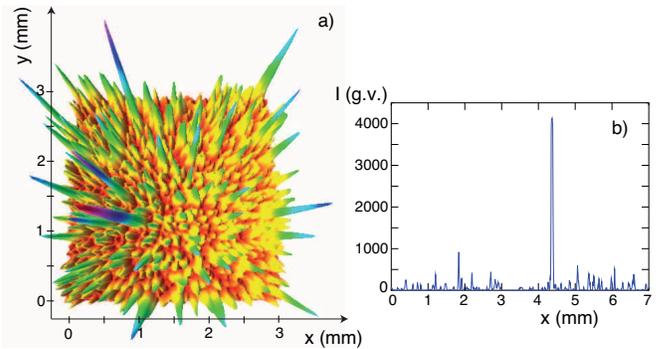}}
\caption{(color online). a) Instantaneous experimental profile of the transverse intensity distribution; b) a 1-D profile showing an extreme event; $I$ is measured in gray values, g.v.}
\label{porcospino}
\end{figure}

The experimental setup is essentially the one described in \cite{ourprl}. The 
ring cavity is formed by three high-reflectivity dielectric mirrors and a lens 
of $f=70$ $cm$ focal length. The total cavity length is $L=273.3$ $cm$ and the 
lens is positioned at a distance $L_1=88.1$ $cm$ from the entrance plane of the 
LCLV. The coordinate system is taken such that $z$ is along the cavity axis 
and $x,y$ are on the transverse plane. 
A LCLV supplies the gain through a two-wave mixing process \cite{IOP} that couples 
the pump beam with the cavity modes. The LCLV is made of a nematic liquid crystal 
layer, thickness $d=14$ $\mu m $, inserted in between a glass wall and a thin slice 
$20$x$30$x$1$ $mm^3$ of the photoconductive $B_{12}SiO_{20}$ (BSO) crystal. A 
voltage $V_0$ is applied by means of transparent electrodes. The working point is 
fixed at $V_0 \simeq 20.3$ $V$, frequency $75$ $Hz$.
The LCLV is pumped by an enlarged and collimated  ($10$ $mm$ diameter) beam from a solid state diode pumped laser ($\lambda=532$ $nm$), linearly polarized in the same direction of the liquid crystal nematic director. The pump and the cavity field, $E_p$ and, respectively, $E_c$, are polarized in the same direction and have a frequency difference of a few $Hz$, the detuning being selected by the voltage $V_0$ applied to the LCLV \cite{ourPRA}.

\begin{figure}[h!]
\centerline{\includegraphics[width=8 cm]{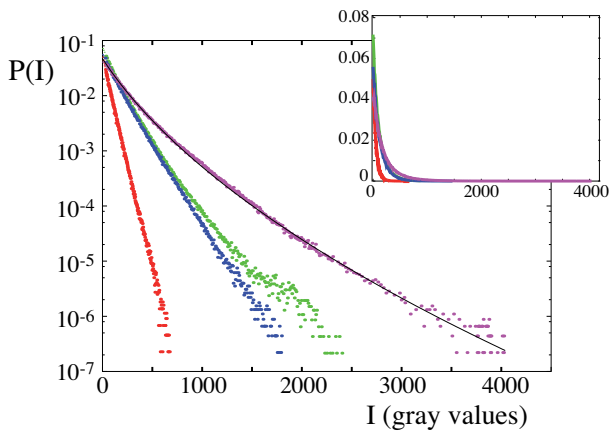}}
\caption{(color online) Experimental PDF of the cavity field intensity; the pump is varied from $I_p/I_{th}=1.8$ (red), $4.0$ (blue), $4.2$ (green) and $6.4$ (purple). 
The red distribution is practically exponential.
The black line is the stretched exponential function that fits the purple distribution, with $c_1=29.5$.
The inset shows the L-shape of the PDF in the linear scale. }
\label{L-stat}
\end{figure}

A small fraction ($4 \%$) of the cavity field is extracted by a beam sampler and sent to a CCD camera ($1024$x$768$ pixels and $12$ bits depth). The detection optical path is set 
so that the camera records the same field distribution as in the entrance plane of the LCLV. 
The Fresnel number is fixed to $F=130$ and can be carefully controlled through a spatial filter introduced near the lens.

For relatively low pump intensity, $I_p= |E_p|^2=2.0$ $mW/cm^2$, the cavity field shows a complex spatio-temporal dynamics, with the formation of many uncorrelated domains. At larger pump, extreme events appear as large amplitude peaks, standing over a speckle-like background. An instantaneous experimental profile of the transverse intensity distribution 
for $I_p= 4.2$ $mW/cm^2$, is shown in Fig.\ref{porcospino}. 
In the inset the 1-D profile shows a large event. Note the difference of the 
peak amplitude with respect to the background values of the intensity.
The locations of the large peaks change spontaneously in time. The typical time-scale of the dynamical evolution is $100$ $ms$, which is ruled by the response time of the liquid crystals \cite{DeGennes}.

The probability density functions (PDF) of the light intensity are determined experimentally by acquiring a large set of images (about one thousand) and then performing the histograms of the intensity values on the whole image stack. In Fig.\ref{L-stat} the PDF of the cavity field intensity, $I=|E_c|^2$, are displayed for different values of the pump, $I_p/I_{th}=1.8$, $4.0$, $4.2$ and $6.4$, $I_{th}=1.2$ $mW/cm^2$ being the threshold for optical oscillations. The dark probability $P(I=0)$ is the same for all
the pump intensities.
The log-linear plots in Fig.\ref{L-stat} reveal increasingly large deviations from the exponential behavior as we increase the pump intensity. All the distributions are well
fitted by the stretched exponential function $N e^{-\sqrt{c_1+c_2 I}}$,
$1/c_1$ providing a measure of the deviation from the exponential function.
The black line is the fitting function of the PDF with $I_p/I_{th}=6.4$.
The fit is performed by setting
the mean intensity and the variance of the fitting function equal to
the experimental values.

Note that an exponential statistics for the intensity corresponds to a Gaussian statistics for the field amplitude. Therefore, an exponential intensity PDF is characteristic of a speckles pattern, where each point receive the uncorrelated contributions of many uncoupled modes. At low pump, this is indeed the behavior displayed by the cavity field. However, when the pump increases, the increasing nonlinear coupling leads to a complex space-time dynamics and extreme events populate the tails of the PDF, providing a large deviation from Gaussianity. The L-shape of the statistics is clearly visible in the inset of Fig.\ref{L-stat}, where the PDF are plotted in linear scale. 

To perform numerical simulations, we have considered the full model equations developed in 
Ref.~\cite{ourPRA}

\begin{eqnarray}
{\partial n_ 0 \over \partial t} &=&-n_0+ \alpha |E_c |^2,
\\
\nonumber
{\partial n_ 1 \over \partial t}&=&-n_1+ \alpha E_c E_p^*,
\end{eqnarray}
where $n_0$ and $n_1$ are, respectively, the amplitude of the homogeneous refractive index and the amplitude of the refractive index grating at the spatial frequency $k_c-k_p$, $k_c$ and $k_p$ being the optical wave numbers of the pump and cavity field. $\alpha$ is the nonlinear coefficient of the LCLV, and we have neglected the diffusion length due to elastic coupling in the liquid crystal. 
The dynamics of the liquid crystals is much slower than the settling of the cavity field, thus $E_c$ follows adiabatically the evolution of $n_0$ and $n_1$.
Then, by taking the wave propagation equation with the cavity boundary conditions, we obtain that
\begin{equation}
E_c=i \sum_{k=0}^\infty{[\hat C e^{i n_0} J_0(2 |n_1|)]^k} \hat C e^{i n_0}{n_1\over |n_1|} J_1(2 |n_1|)E_p,
\end{equation}
where $J_m$ is the Bessel function of the first kind and of order $m$, and $\hat C$ is an operator accounting for the geometry of the cavity and losses, $\hat C= \Gamma^{1/2} e^{i \delta} \hat S_x e^{i (L_0 \nabla_\perp^2 /2 k_p)} e^{-i(k_p \vec r_\perp^2 / 2 f)} e^{i (L_1 \nabla_\perp^2 /2 k_p)}$, with 
$1-\Gamma$ the photon losses, $L_0\equiv L-L_1$ and $\delta$ the phase retardation in a 
round-trip. 
$\hat S_x$ is a symmetry operator that inverts the $x$ axis, thus accounting for 
the odd number of mirrors, $\nabla_\perp^2$ is the transverse Laplacian and 
$\vec r_\perp$ the position in the transverse $(x,y)$ plane.

For the numerical convenience, $\hat C$ has been implemented by using an 
equivalent nearly plane cavity with the $y$-axis inversion $\hat S_y$ and
two lenses just before and after the LCLV~\cite{Siegman}.
In order to account for the finite size of the LCLV and the diaphragm inserted near the lens, 
we have used in the simulations a diaphragm both in the space and in the Fourier plane. 
In particular, we have used a spatial filter with radius equal to $0.4$ $cm$ and a Fourier 
filter with radius $k= 3.2 \cdot10^{-3} k_p$ .
For the other parameters, we have set the liquid crystal response time $\tau=0.1$ $s$, 
the photon loss fraction $1-\Gamma$ between $0.7$ and $0.8$, and $\alpha$ is chosen in 
such a way that the intensity unit is the pump threshold for the activation of 
the cavity field.

\begin{figure}[h!]
\centerline{\includegraphics[width=\columnwidth]{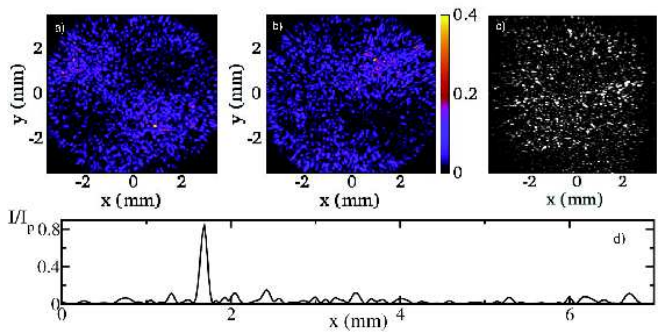}}
\caption{(color online) a) b) Numerical snapshots of the cavity field intensity at 
two different instant times; $I_p=8$, $1-\Gamma= 0.7$; c) experimental snapshot of 
the cavity field, $I_p/I_{th}=4.0$. d) 1D section of a 2D numerical intensity 
distribution.}
\label{numeric}
\end{figure}

Starting with a random initial condition, we observe a transient speckle-like 
behavior, then a breaking of the $y\leftrightarrow-y$ inversion symmetry occurs
and the coarse-grained intensity distribution $|E_c(x,y)|^2$ becomes 
non-homogeneous along the $y$ axis.
In Fig.\ref{numeric} we report two snapshots of the cavity field intensity $|E_c |^2$ 
numerically calculated for $I_p=8$, $1-\Gamma= 0.7$ and displayed at different instant 
times. In Fig.\ref{numeric}a we can note two large dark (bright) regions symmetrically
distributed around the center, while in Fig.\ref{numeric}b the left pair has disappeared 
and the other one is inverted. Typically, the cavity field shows a dynamical evolution 
with the succession of different spatial configurations, each living for a few seconds. 
For comparison, we display in Fig.\ref{numeric}c an experimental snapshot recorded for 
$I_p/I_{th}=4.0$.  
In Fig.3d, we report a one-dimensional section of the
numerical intensity distribution at another instant time,
exhibiting a very large narrow peak in a bright region. Its spatial size is larger
than the experimental value because of a narrower spectral filtering
that limits, for computational convenience, the number of numerical lattice points.

\begin{figure}[h!]
\centerline{\includegraphics[width=8. cm]{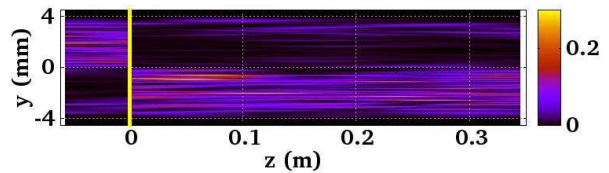}}
\caption{(color online) Cavity field intensity in the $y-z$ plane; the yellow vertical line 
at $z=0$ indicates the position of the LCLV;  $I_p=10$, $1-\Gamma= 0.7$.}
\label{plot}
\end{figure}

In Fig.\ref{plot} we have plotted the 
numerical intensity $|E_c|^2$ in the $z-y$ plane. The yellow line at 
$z=0$ represents the position of the LCLV inside the cavity. 
In the plot, there is clear evidence of both the dynamical symmetry
breaking of the intensity profile along $y$ and of the $y$-axis inversion 
between the impinging and outgoing fields (at left and right of
the yellow line, respectively), consistent with the geometry 
of the experimental cavity, which is nearly spherical and made of an 
odd number of mirrors. This inversion induces a nonlocal coupling 
between different spatial regions of the field, thus triggering the 
hypercycle amplification process that is essential for the
symmetry breaking and the long tail statistics. Indeed, we have checked 
that simulations with a non-inverting nearly plane cavity generate an 
exponential statistics.

\begin{figure}[h!]
\centerline{\includegraphics[width=\columnwidth]{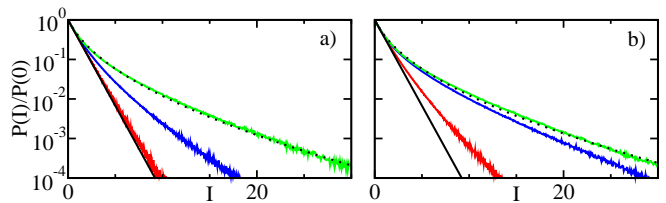}}
\caption{(color online) Numerical PDF of the cavity field intensity for 
$1-\Gamma= 0.7$ (a), $1-\Gamma= 0.8$ (b) and
different values of the pump: $I_p/I_{th}=6$ (blue), $8$ (red), $10$ 
(green). The black straight and dotted lines are respectively $\exp(-I)$ and 
$N\exp(-\sqrt{c_1+c_2 I})$, where $N$ and $c_i$ are set in order to 
fit the green distribution; (a) $c_1=3.73$, (b) $c_1=2.83$.}

\label{pdfnum}
\end{figure}

As a consequence of the spatial symmetry breaking, the field exhibits 
large deviations from Gaussianity. The numerical PDF of the cavity field 
intensity are displayed in Fig.\ref{pdfnum} for different values of the 
pump. The intensities are rescaled in such a way that the PDF have the 
same value and slope at the origin. The distributions are well fitted by 
the stretched exponential function $N \exp(-\sqrt{c_1+c_2 I})$ (black dotted 
lines) and their tails are increasingly populated as the pump increases, 
in agreement with the experimental observations. The simulations show also 
that the symmetry breaking is stronger for high cavity losses.
Moreover, for the used parameters, the refractive index $n_0$ is about four 
times less than $n_1$. The role of $n_0$ is to adjust the global phase 
detuning $\delta$, but it has no effect on the PDF, as we have verified 
by removing it from the model. Thus, in the following 
we take $n_0$ constant and $1- \Gamma \ll1$.

The $y$ axis inversion (Fig.\ref{plot}) 
introduces a nonlocal coupling between different space domains with
loops of amplification, a mechanism similar to the hyper-cycle chain of 
reactions in the catalytic processes \cite{hypercycle}. 
Because of this mechanism, there is a sort of focusing, for which at some space 
locations the cavity field grows much more with respect to the surrounding places, 
giving rise to large amplitude peaks and, hence, to large deviations from the Gaussian 
statistics. The peaks last a few seconds, after that the hyper-cycle readjusts over 
new field configurations, yielding new peaks in other space locations.

To elucidate this mechanism we derive a simple two-mode model, where only the evolution 
of the average refractive index $n_1$ is kept, the average being performed over the 
transverse plane. For a nearly plane cavity, $\bar n_1$ satisfies approximatively
the equation
$\tau {d \bar n_1 \over d t}=-\bar n_1 + I_p F(\bar n_1)$,
where the first and second terms at the r.h.s. account respectively for the liquid 
crystal relaxation and the grating feeding provided by the pump and cavity fields. 
As an ansatz for the function $F$, we take a cubic function, which describes a linear 
growth followed by saturation, due to multiple scattering and pump 
depletion\cite{ourPRA}.

\begin{figure}[h!]
\centerline{\includegraphics[width=8 cm]{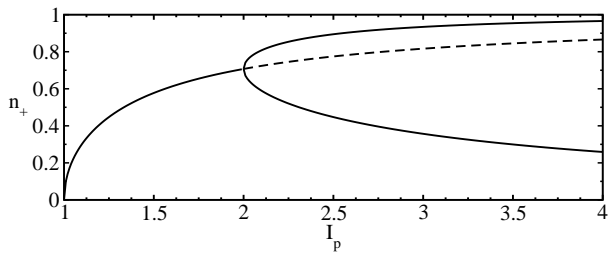}}
\caption{The refractive index $n_+$ as a function of $I_p$.}
\label{diagram}
\end{figure}

When the nearly plane cavity is replaced by a nearly spherical one, the above 
mean-field picture is modified by the nonlocal coupling due to the $y\leftrightarrow -y$
inversion. We thus have to consider two mean fields $\bar n_+$ and $\bar n_-$, 
where the averages are performed over the upper and lower half-planes, respectively. 
Accounting for the inversion $y \rightarrow -y$ of the cavity field, we have that, 
after a round trip, the grating $\bar n_\pm$ is fed by the grating $\bar n_\mp$ 
at the opposite side, i.e, the equation for $\bar n$ is replaced by the two 
following ones for $\bar n_\pm$,
\begin{equation}
\tau {\partial {\bar n}_\pm \over \partial t}=- \bar n_\pm + I_p F(\bar n_\mp),
\label{cycle}
\end{equation}
where we have assumed implicitly that the cavity losses are very high, i.e., the cavity field is negligible after more than one round trip. 

The above equations are similar to a hypercycle model for two autocatalytic systems. 
The main features of $F$ can be captured by the cubic
function $F(\bar n_\pm) =\bar n_\pm - \bar n_\pm^3$. It is easy to show that for $1 < I_p < 2$, 
Eqs.(\ref{cycle}) have only one stable solution, $\bar n_-=\bar n_+=0$, whereas for
$I_p \ge 2$ a bifurcation occurs with the birth of the two stable asymmetric states
$n_+ =2^{-\frac{1}{2}} [1 \pm (I_p^2-4)^{\frac{1}{2}}/I_p]^\frac{1}{2}$,
$n_- =2^{-\frac{1}{2}} [1 \mp (I_p^2-4)^{\frac{1}{2}}/I_p]^\frac{1}{2}$,
which break the $y \leftrightarrow -y$ symmetry, thus qualitatively accounting for the experimental and numerical observations. The bifurcation diagram of $\bar n_+$ is plotted in Fig.\ref{diagram} as function of the pump intensity $I_p$.
The experimental and numerical bifurcation  intensity is about
$4$, rather than $2$. This discrepancy is mainly due to the approximation of
$F(\bar n_\pm)$ with a cubic function.

In conclusion, we have shown experimentally and numerically that extreme waves arise in a spatially extended nonlinear optical system. We have derived a simplified mean-field model, showing that a spatial symmetry breaking is induced by a nonlocal coupling of the cavity field, which is responsible for a hyper-cycle type amplification of the field and leads to a non-Gaussian statistics. 

U.B. and S.R. acknowledge the financial support of the  ANR-07- 
BLAN-0246-03, {\em turbonde}. A.M. and F.T.A. acknowledge the financial support
of Ente Cassa di Risparmio di Firenze under the project "dinamiche cerebrali
caotiche".

\end{document}